\documentclass[showpacs,prb,reprint,floatfix]{revtex4-1}

\pdfoutput=1
\usepackage{graphicx,psfrag,epsf,epsfig}
\usepackage{pstricks,pst-plot}      
\usepackage{latexsym}
\usepackage{mathrsfs,amsmath,amssymb,textcomp}
 \usepackage{flushend}
\usepackage{hyperref}
\usepackage{feynmf}

\def\d{{\mathrm d}}
\newcommand{\pf}[1]{\frac{\partial}{\partial #1}}

\begin{document}


\title{Quantum-kinetic theory of steady-state photocurrent generation in thin films: Coherent versus
incoherent coupling}

\author{U. Aeberhard}
\email{u.aeberhard@fz-juelich.de}

\affiliation{IEK-5: Photovoltaik, Forschungszentrum J\"ulich, D-52425 J\"ulich, Germany}

\date{\today}

\begin{abstract}
The generation of photocurrents due to coupling of electrons to both classical and
quantized electromagnetic fields in thin semiconductor films is described within the framework of the nonequilibrium Green's 
function formalism. For the coherent coupling to classical fields corresponding to single field
operator averages, an effective two-time intraband self-energy is derived from a band decoupling procedure.
The evaluation of coherent photogeneration is performed self-consistently with the
propagation of the 	fields by using for the latter a transfer matrix formalism with an extinction
coefficient derived from the electronic Green's functions. For the ``incoherent'' coupling to fluctuations
of the quantized fields, which need to be considered for the inclusion of spontaneous emission, the
first self-consistent Born self-energy is used, with full spatial resolution in the photon Green's
functions. These are obtained from the numerical solution of Dyson and Keldysh equations including a
nonlocal photon self-energy based on the same interband polarization function as used for the
coherent case. 	A comparison of the spectral and integral photocurrent generation pattern reveals a
close agreement between coherent and incoherent coupling for the case of an ultra-thin, selectively
contacted absorber layer at short circuit conditions.
\end{abstract}

\maketitle

\section{\label{sec:level1}Introduction} 

Among the state-of-the-art theories used to describe the operation of
complex nanostructure-based optoelectronic devices, e.g., quantum well and quantum dot lasers and
light-emitting diodes, quantum-kinetic formalisms are most powerful in terms of both physical insight and predictive
power provided \cite{haug:04}. However, the simulation of devices such as nanostructure-based solar cells requires  the
development of a unified picture of quantum optics and quantum transport \cite{ae:jcel_11}, since an accurate
description of both optics and charge transport is crucial to capture the impact of complex dielectric and
electronic nanostructure potentials on the device performance \cite{ae:jstqe_13}.
There, one is faced with the problem of different representations of the quantum-kinetic theory conventionally used. In
quantum optics, the focus is on transient or ultrafast phenomena, with standard descriptions based on density matrix 
theory corresponding to the equal time Green's function formalism \cite{schaefer:02,kira:06}. In quantum transport, the
operating regime of interest is the steady state, which on the quantum-kinetic level is described by using the Fourier transform to
the energy domain of the relative time in the two-time Green's function \cite{datta:95,datta:05}.   

In this paper, two different approaches to the solution of the problem are presented. Both are concerned with the
formulation of an electron-photon self-energy compatible with the steady-state non-equilibrium Green's function
(NEGF) formalism of quantum transport. In the first case, the self-energy describes the coupling of the electronic
system to coherent fields as obtained from classical solutions of Maxwell's equations. It establishes the
connection to conventional description of light attenuation in solar cells and provides a computationally efficient
treatment of stimulated electron-photon processes within the NEGF formalism. The second type of
electron-photon self-energy relates to the non-local photon Green's function (GF) on the level of quantum statistical
mechanics and includes the coupling to incoherent field fluctuations. It enables a consistent description of optical generation and radiative
recombination by including the coupling to any available photon modes of the device,  from the leaky modes occupied by
incident solar photons to the guided modes populated by spontaneous emission. This is an essential prerequisite for the
assessment of the radiative efficiency limit in novel nanostructure-based solar cell devices.

The paper is organized as follows. After the formulation of the general NEGF theory of
optoelectronic processes for a two-band semiconductor model in Sec.~\ref{sec:2band}, the effective
self-energy for coherent coupling is derived in Sec.~\ref{sec:coherent_se}. The main body of the 
paper is formed by Sec.~\ref{sec:photogen} on the application of the NEGF theory to the simulation
of charge carrier photogeneration in thin semiconductor films for coupling to classical, average
fields and to the non-equilibrium statistical ensemble average of field operator pairs. For both classical and quantized
fields, expressions for the local photogeneration rate, the local absorption coefficient and the absorptance, as well as
the resulting photocurrent, are formulated in the microscopic NEGF picture and evaluated via numerical simulation for a
prototypical thin-film solar cell architecture.

\section{\label{sec:2band} Quantum-kinetic theory of electron-photon interaction in a two band
model}

The electronic model system under investigation is a simple two band model of a direct gap
semiconductor film furnished with ohmic contacts
and coupled to an external photon field\cite{ae:prb_08}, which is treated either classically or
quantum mechanically. The Hamiltonian of the electronic system thus reads
\begin{align}
\hat{H}=\hat{H}_{0}+\hat{H}_{e\gamma}+\hat{H}_{C}+\hat{H}_{diss}.\label{eq:hamiltonian}
\end{align}
$\hat{H}_{0}$ is the Hamiltonian of the noninteracting isolated
mesoscopic absorber plus the Hartree term $U(\mathbf{r})$ from
the solution of the macroscopic Poisson equation, corresponding to
a mean-field treatment of carrier-carrier interaction. $\hat{H}_{e\gamma}$ describes the
light-matter interaction and $\hat{H}_{C}$ the (selective) coupling to contacts. At this level,
the theory is able to cover the operation of an ideal solar cell. The generic term $\hat{H}_{diss}$
encodes dissipative processes such as electron-phonon interaction as well as non-radiative
recombination processes and is not considered here. The carriers in conduction ($c$) and valence
($v$) bands are described by field operators $\hat{\Psi}_{b}(\mathbf{r},t)$, $b=c,v$, defining the
charge carrier Green's functions via
\begin{equation}
G_{ab}(\underline{1},\underline{1}')\equiv -\frac{i}{\hbar}\langle
\hat{\Psi}_{a}(\underline{1})\hat{\Psi}_{b}^{\dagger}(\underline{1}')\rangle_{\mathcal{C}}\label{eq:carr_gf}
\end{equation}
for band indices $\{a,b\}\in\{c,v\}$ and $\underline{1}\equiv(\mathbf{r}_{1},\underbar{t}_{1}\in
\mathcal{C})$, where $\mathcal{C}$ denotes the Keldysh contour\cite{keldysh:65}. The electromagnetic field is
described in terms of the vector potential operator,
$\hat{\mathbf{A}}=\hat{\mathbf{A}}_{coh}+\hat{\mathbf{A}}_{inc}$ which is 	decomposed into a
coherent contribution $\hat{\mathbf{A}}_{coh}$ corresponding to coherent light sources and a 
contribution $\hat{\mathbf{A}}_{inc}$ from incoherent light sources and from spontaneous emission.
The coherent vector potential is related to the time-dependent part of the classical electric field
$\boldsymbol{\mathcal{E}}$ 	via the standard relation
\begin{align}
-\pf{t}\langle\hat{\mathbf{A}}_{coh}(\mathbf{r},\underline{t})\rangle_{\mathcal{C}}=\boldsymbol{\mathcal{E}}(\mathbf{r},t).
\end{align}
The photon Green's function $\mathcal{D}$, on the other hand, includes the incoherent field
fluctuations\cite{richter:08},
\begin{align}
\mathcal{D}_{\mu\nu}(\underbar{1},\underbar{2})=&
-\frac{i}{\mu_{0}\hbar}\left[\langle \hat{A}_{\mu}(\underbar{1})
\hat{A}_{\nu}(\underbar{2})\rangle_{\mathcal{C}}-\langle \hat{A}_{\mu}(\underbar{1})\rangle_{\mathcal{C}}\langle
\hat{A}_{\nu}(\underbar{2})\rangle_{\mathcal{C}}\right],\label{eq:photgf}
\end{align}
where $\mu_{0}$ is the magnetic vacuum permeability. 

In terms of the above field operators, the electron-photon coupling component in the Hamiltonian for
the electronic system is expressed as follows:
\begin{align}
 [\mathcal{H}_{e\gamma}]_{ab}(t)=&\int d^{3}r
 \hat{\Psi}_{a}^{\dagger}(\mathbf{r},t)\hat{H}_{e\gamma}(\mathbf{r},t)\hat{\Psi}_{b}(\mathbf{r},t),
 \label{eq:elphotham}\\
 \hat{H}_{e\gamma}(\mathbf{r},t)=&-\frac{e}{m_{0}}\hat{\mathbf{A}}(\mathbf{r},t)\cdot\hat{\mathbf{p}}.
\end{align}
In the NEGF picture of photogeneration, the effect of irradiation on the electronic system is
considered in the form of a self-energy that renormalizes the carrier Green's function in the
solution of the Dyson equations. For weak coupling, the self-energy can be derived within
many-body perturbation theory using the Hamiltonian \eqref{eq:elphotham}, from the expansion of the
perturbed (photon assisted) carrier Green's function 
\begin{align}
G(\mathbf{r},t;\mathbf{r}',t')= -\frac{i}{\hbar}\left\langle
e^{-\frac{i}{\hbar}\int_{\mathcal{C}}ds
\mathcal{\hat{H}}_{e\gamma}(s)}\hat{\Psi}(\mathbf{r},\underline{t})\hat{\Psi}^{\dagger}(\mathbf{r}',\underline{t}')\right\rangle_{\mathcal{C}}.\label{eq:pertexp_el}
\end{align}
To first order in the vector potential, perturbation theory results in
the singular self-energy term
\begin{align} 
\Sigma_{e\gamma}^{\delta}(\underline{1})=-\frac{e}{m_{0}}\langle
\hat{\mathbf{A}}(\underline{1})\rangle_{\mathcal{C}}\cdot \hat{\mathbf{p}}(1)\equiv-\frac{
e}{m_{0}}\hat{\mathbf{A}}_{coh}(\underline{1})\cdot \hat{\mathbf{p}}(1),\label{eq:singular_se}
\end{align}
which corresponds to the interband term in the effective Hamiltonian originating
in the coupling to transverse photons. To second order in the vector
potential, the self-energy corresponds to the random phase approximation (RPA)
expression\cite{pereira:96,pereira:98}
\begin{align}
\Sigma^{RPA}(\underbar{1},\underbar{2})=&i\hbar\mu_{0}\Big(\frac{e}{m_{0}}\Big)^{2}\sum_{\mu\nu}\hat{p}^{\mu}(1,1')
G(\underbar{1},\underbar{2})\nonumber\\&\times\hat{p}^{\nu}(2)\mathcal{D}_{\mu\nu}(\underbar{2},\underbar{1}')|_{1'=1},
\label{eq:elphotse_rpa}
\end{align}
where $\hat{p}$ is the momentum operator and $\hat{p}^{\mu}(1,1')\equiv
[\hat{p}^{\mu}(1)-\hat{p}^{\mu}(1')]/2$. This self-energy can also be derived via functional
derivative techniques\cite{dubois:67,henneberger:88_3,jahnke:95}. The photon Green's functions in
\eqref{eq:elphotse_rpa} are obtained from the corresponding Dyson
equations\cite{jahnke:95,henneberger:96}:
\begin{align}
\int_{\mathcal{C}}
d3\left[\mathcal{D}_{0,\mu\nu}^{-1}(\underbar{1},\underbar{3})-\Pi_{\mu\beta}(\underbar{1},\underbar{3})
\right]\mathcal{D}_{\beta\nu}(\underbar{3},\underbar{2})
=\delta_{\parallel,\mu\nu}(\underbar{1},\underbar{2}),\label{eq:contour_dyson_phot}
\end{align}
where $\mathcal{D}_{0,\mu\nu}$ is the free propagator defined by 
\begin{align}
\mathcal{D}_{0,\mu\nu}^{-1}(\underbar{1},\underbar{2})=\left[\Delta_{1}-\frac{1}{c^2}\frac{\partial^2}{\partial
t_{1}^2}\right]\delta_{\mu\nu}\delta(\underbar{1},\underbar{2}),
\end{align}
and $
\delta_{\parallel,\mu\nu}(\underbar{1},\underbar{2})=\delta(\underline{t}_{1}-\underline{t}_{2})\delta_{\parallel,\mu\nu}(\mathbf{r}_1-\mathbf{r}_2)$
is the transverse $\delta$ function. In Eq. \eqref{eq:contour_dyson_phot}, $\boldsymbol{\Pi}$ is the
photon self-energy describing the renormalization of the photon Green's function due to interaction with
the electronic system. 	The RPA photon self energy due to interband transitions	\footnote{Intraband
transitions such as free carrier absorption are explicitely excluded at this stage.} is given in terms 	of electronic Green's
functions and momentum operator elements as follows\cite{pereira:96,pereira:98}:
\begin{align}
\Pi_{\mu\nu}^{RPA}(\underbar{1},\underbar{2})=&-i\hbar\mu_{0}\Big(\frac{e}{m_{0}}\Big)^{2}\hat{p}^{\mu}(1,1')
G(\underbar{1},\underbar{2})\nonumber\\&\times\hat{p}^{\nu}(2)G(\underbar{2},\underbar{1}')|_{1'=1}.\label{eq:photse_rpa}
\end{align}

For the comparison of the two types of electron-photon self-energies, the effects of the singular
interband term needs to be transferred to an effective two-time intraband self-energy, which can be
achieved via a band decoupling procedure, as shown below.

\section{\label{sec:coherent_se}Effective interband self-energy for coherent coupling}

In the following, a band decoupling scheme similar to that
introduced in Ref.~\onlinecite{henneberger:88} will be applied to the two band model. The procedure
was given in the appendix of Ref.~\onlinecite{ae:prb_11} for a general singular self-energy term,
but for the sake of clarity and completeness will be repeated here for the specific case of the electron-photon
interaction.\footnote{If the singular part of the electron-electron interaction is
included, the singular self-energy is no longer local in space and spatial integration should also
be performed.} Starting point are the Kadanoff-Baym equations for contour-ordered
non-equilibrium Green's functions\cite{kadanoff:62},
\begin{align}
\mathbf{G}_{0}^{-1}(\underline{1},\underline{1})\mathbf{G}(\underline{1},\underline{1}')=&
\boldsymbol{\delta}(\underline{1},\underline{1}') +\int_{\mathcal{C}}\d2\,\boldsymbol{\Sigma}(\underline{1},\underline{2})\mathbf{G}(\underline{2},\underline{1}'),\label{eq:diffdys1}\\
[\mathbf{G}_{0}^{\dagger}]^{-1}(\underline{1}',\underline{1}')\mathbf{G}(\underline{1},\underline{1}')=
&\boldsymbol{\delta}(\underline{1},\underline{1}')+\int_{\mathcal{C}}\d 2
\,\mathbf{G}(\underline{1},\underline{2})
\boldsymbol{\Sigma}(\underline{2},\underline{1}'),\label{eq:diffdys2}
\end{align} 
where 
\begin{equation}
[G_{0}^{-1}(\underline{1},\underline{1}')]_{ab}=\left(i\hbar\pf{t_{1}}-[H_{0}(\mathbf{r}_{1})]_{a}\right)\delta(\underline{1},
\underline{1}')\delta_{ab},
\end{equation}
and $G_{ab}$ is defined in \eqref{eq:carr_gf}. Real-time decomposition rules
\cite{langreth:76} applied to \eqref{eq:diffdys1} provide the coupled equations
for the retarded components of the intra- and interband Green's functions, 
\begin{align}
G^{-1}_{0,cc}(1,1)G^{R}_{cc}(1,1')=&\delta(1,1')
+\Sigma_{cv}^{\delta}(1) G_{vc}^{R}(1,1')\nonumber\\&+\int\d
2\,\Sigma_{cc}^{R}(1,2)G^{R}_{cc}(2,1'),\label{eq:GRcc}\\
G^{-1}_{0,vv}(1,1)G^{R}_{vc}(1,1')=&\Sigma_{vc}^{\delta}(1)
G_{cc}^{R}(1,1')\nonumber\\&+\int\d
2\,\Sigma_{vv}^{R}(1,2)G^{R}_{vc}(2,1').\label{eq:GRvc}
\end{align}
Introducing the new quantity 
\begin{align}
\tilde{G}_{vv}^R\equiv\left[G^{-1}_{0,vv}-\Sigma_{vv}^{R}\right]^{-1} 
\end{align}
in \eqref{eq:GRvc}, the retarded interband GF can be written as
\begin{align}
G_{vc}^{R}(1,1')=\int\d
2\,\tilde{G}_{vv}^R(1,2)\Sigma_{vc}^{\delta}(2)
G_{cc}^{R}(2,1').
\end{align}
Inserting the above expression in \eqref{eq:GRcc} yields a closed equation for
the intraband GF,
\begin{align}
G_{cc}^{R}(1,1')=&\Big[G^{-1}_{0,cc}(1,1')-\Sigma_{cc}^{R}(1,1')
\nonumber\\&-\Sigma_{cv}^{\delta}(1)\tilde{G}_{vv}^{R}(1,1')
\Sigma_{vc}^{\delta}(1')\Big]^{-1}\\
 \equiv&\left[\tilde{G}^{-1}_{cc}(1,1')-\tilde{\Sigma}_{cc}^{R}(1,1')\right]^{-1},\label{eq:dyson_gr}
\end{align}
where the effective band-coupling self-energy $\tilde{\Sigma}$ was defined,
\begin{align}
\tilde{\Sigma}_{cc}^{R}(1,1')\equiv\Sigma_{cv}^{\delta}(1)\tilde{G}_{vv}^R(1,1')
\Sigma_{vc}^{\delta}(1').
\end{align}
Similarly, the lesser and greater components of the Green's functions can
be decoupled: starting from
\begin{align}
&G^{-1}_{0,cc}(1,1)G^{<}_{cc}(1,1')=\Sigma_{cv}^{\delta}(1)
G_{vc}^{<}(1,1')\nonumber\\&+\int\d
2\,\Sigma_{cc}^{R}(1,2)G^{<}_{cc}(2,1')+\int\d
2\,\Sigma_{cc}^{<}(1,2)G^{A}_{cc}(2,1'),\label{eq:Gincc}\\
&G^{-1}_{0,vv}(1,1)G^{<}_{vc}(1,1')=\Sigma_{vc}^{\delta}(1)
G_{cc}^{<}(1,1')\nonumber\\&+\int\d
2\,\Sigma_{vv}^{R}(1,2)G^{<}_{vc}(2,1')+\int\d
2\,\Sigma_{vv}^{<}(1,2)G^{A}_{vc}(2,1'),\label{eq:GRinvc}
\end{align}
the interband correlation or coherent polarization function is written as
\begin{align}
G_{vc}^{<}(1,1')=&\int\d
2\,\Big[\tilde{G}_{vv}^{R}(1,2)\Sigma_{vc}^{\delta}(2)
G_{cc}^{<}(2,1')\nonumber\\&
+\tilde{G}_{vv}^{<}(1,2)\Sigma_{vc}^{\delta}(2)	
G_{cc}^{A}(2,1')\Big],
\end{align}
where 
\begin{align}
\tilde{G}_{vv}^{<}(1,1')\equiv\int\d
2\,\int\d 3\,
\tilde{G}_{vv}^{R}(1,2)\Sigma_{vv}^{<}(2,3)
\tilde{G}_{vv}^{A}(3,1')
\end{align}
was introduced. Replacing the interband term in \eqref{eq:Gincc} then yields
the intraband correlation function 
\begin{align}
G_{cc}^{<}(1,1')=\int\d
2\,\int\d
3\,&
G_{cc}^{R}(1,2)\Big[\Sigma_{cc}^{<}(2,3)\nonumber\\&
+\tilde{\Sigma}_{cc}^{<}(2,3)\Big]G_{cc}^{A}(3,1')\label{eq:keldysh_gn}
\end{align}
with
\begin{align}
\tilde{\Sigma}_{cc}^{<}(1,1')\equiv
\Sigma_{cv}^{\delta}(1)\tilde{G}_{vv}^{<}(1,1')
\Sigma_{vc}^{\delta}(1').
\end{align}
The expressions for the valence band self-energy corrections are
obtained from analogous derivations as 
\begin{align}
\tilde{\Sigma}_{vv}^{\alpha}(1,1')=&\Sigma_{cv}^{\delta}(1)
\tilde{G}_{cc}^{\alpha}(1,1')\Sigma_{cv}^{\delta}(1'),\quad
\alpha=R,A,\lessgtr.
\end{align}

We can now evaluate the effective band-coupling self-energy expression using the
singular interband self-energies due to the coupling to coherent radiation or
classical fields given in the previous section [Eq. \eqref{eq:singular_se}] and compare it
to the self-energy for the coupling to incoherent radiation [Eq.\eqref{eq:elphotse_rpa}]. In
steady-state conditions, the dependence on microscopic time vanishes and the relative time
dependence is Fourier transformed to the energy domain. As shown in
Ref.~\onlinecite{henneberger:88_3}, using the rotating wave approximation, the coherently driven
interband self-energies have a combined time-dependence $\propto e^{\pm i\omega \tau}$, where
the sign depends on the band index and $\tau=t_{1}-t_{2}$ is the relative time, which provides
the Fourier transform
\begin{align}
&\tilde{\Sigma}_{cc}^{\lessgtr}(\mathbf{r}_{1},\mathbf{r}_{2},E)=\int d\hbar\omega\Big[
\Sigma_{cv}^{\delta}(\mathbf{r}_{1},\hbar\omega)\nonumber\\&\times\int
d\tau
\tilde{G}_{vv}^{\lessgtr}(\mathbf{r}_{1},\mathbf{r}_{2},\tau)e^{\frac{i}{\hbar}(E-\hbar\omega)\tau}
\Sigma_{vc}^{\delta}(\mathbf{r}_{2},\hbar\omega)\Big]\\ 
&=\int d\hbar\omega \Sigma_{cv}^{\delta}(\mathbf{r}_{1},\hbar\omega)
\tilde{G}_{vv}^{\lessgtr}(\mathbf{r}_{1},\mathbf{r}_{2},E-\hbar\omega)
\Sigma_{vc}^{\delta}(\mathbf{r}_{2},\hbar\omega),
\end{align}
and
\begin{align}
\tilde{\Sigma}_{vv}^{\lessgtr}(\mathbf{r}_{1},\mathbf{r}_{2},E)=&\int
d\hbar\omega\Sigma_{vc}^{\delta}(\mathbf{r}_{1},\hbar\omega)\nonumber\\&\times
\tilde{G}_{cc}^{\lessgtr}(\mathbf{r}_{1},\mathbf{r}_{2},E+\hbar\omega)
\Sigma_{cv}^{\delta}(\mathbf{r}_{2},\hbar\omega).
\end{align}
Inserting the explicit form of the coherent singular self-energies \eqref{eq:singular_se} and
integrating over photon energies $E_{\gamma}=\hbar\omega$ yields the general polychromatic expression
\begin{align}
&\tilde{\Sigma}_{aa}^{\lessgtr}(\mathbf{r}_{1},\mathbf{r}_{2},E)=\Big(\frac{e}{m_{0}}\Big)^{2}\sum_{\mu\nu}\int
dE_{\gamma} A_{\mu}(\mathbf{r}_{1},E_{\gamma})
p_{ab}^{\mu}(\mathbf{r}_{1})\nonumber\\&\qquad\times\tilde{G}_{bb}^{\alpha}(\mathbf{r}_{1},\mathbf{r}_{2},E\mp
E_{\gamma})A_{\nu}^{*}(\mathbf{r}_{2},E_{\gamma})p_{ab}^{\nu*}(\mathbf{r}_{2}),\label{eq:se_coh}
\end{align}
where the negative sign is for $a=c$ and the electromagnetic vector potential is normalized to the
total intensity, i.e., it has units (eV)$^{\frac{1}{2}}$/(Am).

\section{\label{sec:photogen} Photogeneration in thin films}

\begin{figure}[t]
\includegraphics[width=0.4\textwidth]{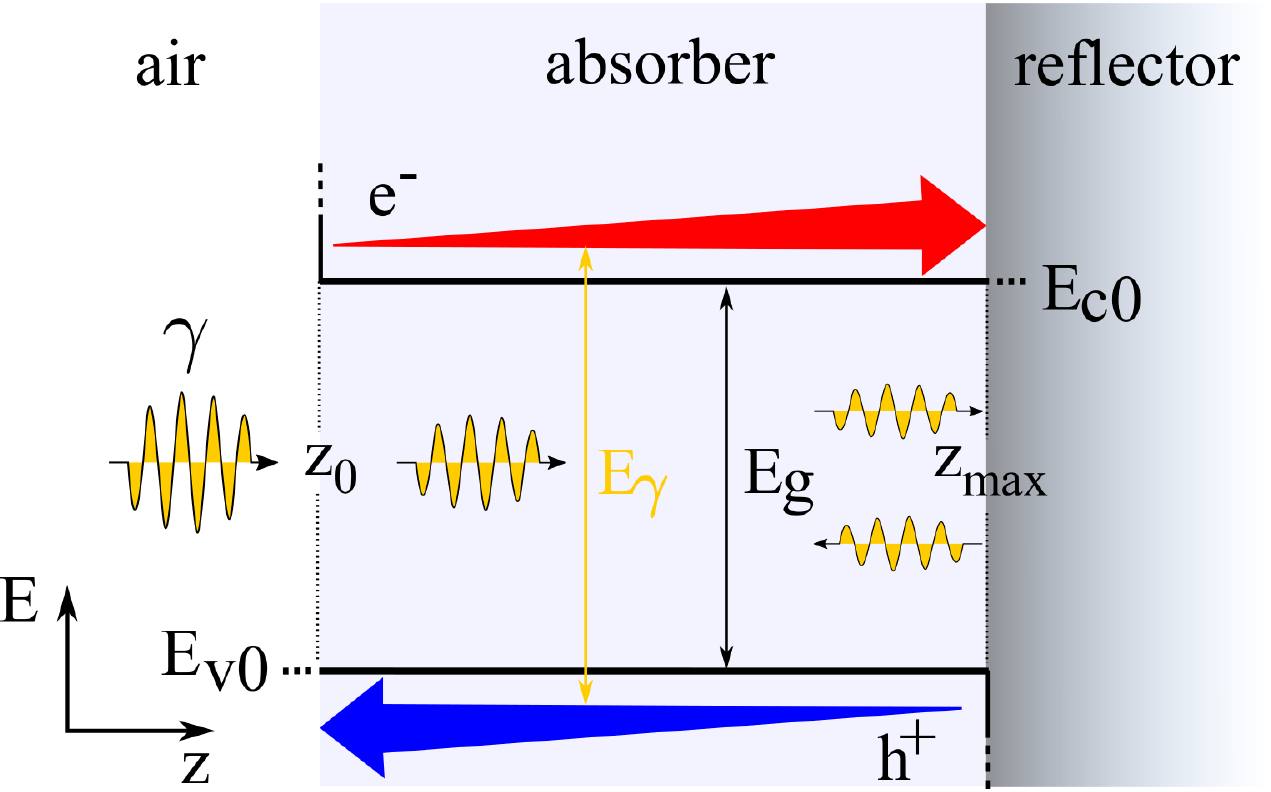} 
\caption{(Color online) Schematic band diagram representation of a flat band solar cell architecture
with finite photocurrent flow enabled by carrier selective contacts: electrons ($e^{-}$) are blocked in
the conduction band at $z=z_{0}$ via an infinite barrier potential in the conduction band energy
$E_{C}$, while holes ($h^{+}$) are reflected at $z=z_{max}$ due to a similar barrier potential in
$E_{V}$. For closer resemblance to actual thin-film solar cell configurations, an additional back
reflector layer is considered in the optical simulations.
\label{fig:structure}}
\end{figure}

A large variety of advanced optoelectronic devices with nanoscale active region are based on
ultra-thin semiconductor films. Here, as a simple model system for the evaluation of electron-photon coupling,
an intrinsic GaAs slab at short circuit conditions is used. Photocurrent rectification is achieved via the imposition of
carrier selective contacts, which is sufficient for photovoltaic operation \cite{wuerfel:book_05},
and has already been applied in the NEGF simulation of nanostructured solar cells \cite{ae:oqel_12}. 
The architecture shown in Fig.~\ref{fig:structure} deviates minimally from the standard flat band
bulk situation in terms of electronic structure, but provides at the same time complete charge
separation. For the system to show more
resemblance to the situation encountered in actual thin-film solar cell devices, a silver back 
reflector contact layer is added to the right of the slab for the
optical simulation.

Before comparing the optoelectronic response of a given structure to classical and quantized
fields, the representation of the NEGF formalism is adjusted to the slab system at hand.
\newline

\subsection{Electronic and optical states in thin semiconductor films}
In the layer structure with homogeneous transverse dimensions,
the description of the optoelectronic properties can be simplified by using a plane wave expansion of the field operators for electrons and photons with respect to transverse coordinates,
i.e.,
\begin{align}
\hat{\Psi}(\mathbf{r},t)=&\mathcal{A}^{-\frac{1}{2}}\sum_{\mathbf{k}_{\parallel}}\hat{\Psi}(\mathbf{k}_{\parallel},z,t)
e^{i\mathbf{k}_{\parallel}\cdot\mathbf{r}_{\parallel}},\\
\hat{\mathbf{A}}(\mathbf{r},t)=&\mathcal{A}^{-\frac{1}{2}}\sum_{\mathbf{q}_{\parallel}}\hat{\mathbf{A}}(\mathbf{q}_{\parallel},z,t)
e^{i\mathbf{q}_{\parallel}\cdot\mathbf{r}_{\parallel}},
\end{align}
where $\mathcal{A}$ denotes the cross section area of the film. The corresponding slab representation for the steady-state Green's functions is obtained from
\begin{align}
G(\mathbf{r},\mathbf{r}',E)=&\mathcal{A}^{-1}\sum_{\mathbf{k}_{\parallel}}G(\mathbf{k}_{\parallel},z,z',E)
e^{i\mathbf{k}_{\parallel}\cdot(\mathbf{r}_{\parallel}-\mathbf{r}'_{\parallel})},\\
\mathcal{D}_{\mu\nu}(\mathbf{r},\mathbf{r}',E)=&\mathcal{A}^{-1}\sum_{\mathbf{q}_{\parallel}}\mathcal{D}_{\mu\nu}(\mathbf{q}_{\parallel},z,z',E)
e^{i\mathbf{q}_{\parallel}\cdot(\mathbf{r}_{\parallel}-\mathbf{r}'_{\parallel})}.\label{eq:photongf_slab}
\end{align}
For each energy and transverse momentum vector, a separate set of equations for the Green's
functions needs to be solved. For the charge carriers, the steady-state
integro-differential equations derived from \eqref{eq:diffdys1} and \eqref{eq:diffdys2} read 
\begin{widetext}	
\begin{align}
G^{R}(\mathbf{k}_{\parallel},z,z',E)=&G_{0}^{R}(\mathbf{k}_{\parallel},z,z',E)+\int
dz_{1}\int
dz_{2}~G_{0}^{R}(\mathbf{k}_{\parallel},z,z_{1},E)\Sigma^{R}(\mathbf{k}_{\parallel},z_{1},z_{2},E)
G^{R}(\mathbf{k}_{\parallel},z_{2},z',E),\\
G^{\lessgtr}(\mathbf{k}_{\parallel},z,z',E)=&\int
dz_{1}\int
dz_{2}~G^{R}(\mathbf{k}_{\parallel},z,z_{1},E)\Sigma^{\lessgtr}(\mathbf{k}_{\parallel},z_{1},z_{2},E)
G^{A}(\mathbf{k}_{\parallel},z_{2},z',E),
\end{align}
\end{widetext}
with
\begin{align}
\big[E-\mathcal{\hat{H}}_{0}(\mathbf{k}_{\parallel},z)\big]G_{0}^{R}(\mathbf{k}_{\parallel},z,z',E)=\delta(z-z').
\end{align}
The numerical evaluation of the above equations employs a real-space basis in the
dimension perpendicular to the film, and a plane-wave expansion in the in-plane dimensions
 in combination with a two-band effective mass Hamiltonian for the electronic structure
 \cite{ae:nrl_11}. 
 
 The Dyson and Keldysh equations for the dyadic photon Green's functions are found
 in analogy to their electronic counterparts in the following form (assuming again
 summation over repeated polarization indices):
 \begin{widetext}
\begin{align}
\mathcal{D}_{\mu\nu}^{R}(\mathbf{q}_{\parallel},z,z',E)=&\mathcal{D}_{0\mu\nu}^{R}(\mathbf{q}_{\parallel},z,z',E)
+\int
dz_{1}\int dz_{2}\mathcal{D}_{0\mu\alpha}^{R}(\mathbf{q}_{\parallel},z,z_{1},E)\Pi_{\alpha\beta}^{R}(\mathbf{q}_{\parallel},z_{1},z_{2},E)
\mathcal{D}_{\beta\nu}^{R}(\mathbf{q}_{\parallel},z_{2},z',E),\\
\mathcal{D}_{\mu\nu}^{\lessgtr}(\mathbf{q}_{\parallel},z,z',E)=&\int dz_{1}\int
dz_{2}\mathcal{D}_{\mu\alpha}^{R}(\mathbf{q}_{\parallel},z,z_{1},E)\Big[
\Pi_{0\alpha\beta}^{\lessgtr}(\mathbf{q}_{\parallel},z_{1},z_{2},E)+
\Pi_{\alpha\beta}^{\lessgtr}(\mathbf{q}_{\parallel},z_{1},z_{2},E) \Big]
\mathcal{D}_{\beta\nu}^{A}(\mathbf{q}_{\parallel},z_{2},z',E),\label{eq:keldysh}
\end{align}
\end{widetext}
where the self-energy component related to the solution of the \emph{homogeneous} problem, i.e.,
incident fluctuations that are independent from the state of the absorber, is given
by\cite{mozyrsky:07,richter:08}
\begin{align}
&\Pi_{0\mu\nu}^{\lessgtr}(\mathbf{q}_{\parallel},z,z',E)=\int dz_{1}\int
dz_{2}[\mathcal{D}_{0}^{R}]^{-1}_{\mu\alpha}(\mathbf{q}_{\parallel},z,z_{1},E)\nonumber\\\qquad&\times
\mathcal{D}_{0\alpha\beta}^{\lessgtr}(\mathbf{q}_{\parallel},z_{1},z_{2},E) 
[\mathcal{D}_{0}^{A}]^{-1}_{\beta\nu}(\mathbf{q}_{\parallel},z_{2},z',E)
\end{align}
in terms of the Green's functions of the unperturbed system. These equations are solved in real
space using a numerical quadrature method\cite{ae:photgf}. 

The classical fields for the evaluation of the coherent self-energy \eqref{eq:se_coh} are computed
using a conventional transfer matrix method (TMM), with extinction coefficient obtained from the
absorption coefficient computed within the NEGF formalism, i.e., in complete consistency with the transport
properties, such as the photocurrent generated via coupling to the EM field, as shown in the
following.

\subsection{Absorption and photogeneration}
The local absorption coefficient at a fixed energy ($\sim E_{\gamma}$), polarization ($\sim \mu$) and angle of incidence ($\sim
\mathbf{q}_{\parallel},~E_{\gamma}$) is related to the corresponding local and spectral photogeneration rate $g$ (per unit volume)
and local photon flux $\Phi$ (per unit photon energy) via
\begin{align}
g^{\mu}(\mathbf{q}_{\parallel},z,E_{\gamma})=\Phi_{\mu}(\mathbf{q}_{\parallel},z,E_{\gamma})	
\alpha_{\mu}(\mathbf{q}_{\parallel},z,E_{\gamma}).\label{eq:genrate_def}
\end{align}
The spectral photogeneration rate can be obtained from the expression for the local integral
radiative interband volume generation rate $\mathcal{G}$ in terms of electronic Green's functions and
self-energies \cite{ae:prb_11}, which for charge carriers in the CB reads
\begin{align}
\mathcal{G}_{c}(z)=&\mathcal{A}^{-1}\sum_{\mathbf{k}_{\parallel}}\int dz'
\int\frac{dE}{2\pi\hbar}\Sigma_{cc}^{<}(\mathbf{k}_{\parallel},z,z',E)\nonumber\\&\times
G_{cc}^{>}(\mathbf{k}_{\parallel},z',z,E)\\
\equiv& \mathcal{A}^{-1}\sum_{\mu}\sum_{\mathbf{q}_{\parallel}}\int
dE_{\gamma} ~g^{\mu}_{c}(\mathbf{q}_{\parallel},z,E_{\gamma}).\label{eq:integral_genrate}
\end{align}
In principle, the self-energy term contains all the scattering mechanisms present in the interaction part of the Hamiltonian \eqref{eq:hamiltonian}. While the direct contribution of intraband scattering vanishes upon energy integration over the band,  dissipative intraband processes, such as electron-phonon interaction,  still affect the rate via the dressing of the full GF in \eqref{eq:integral_genrate}. At the radiative limit, the
photocurrent at zero bias voltage, i.e., the short circuit current density $J_{sc}$, is directly
obtained from the incident photon flux and the total absorptance of the slab,
\begin{align}
J_{sc}=\frac{e}{\mathcal{A}}\sum_{\mathbf{q}_{\parallel}}\int
dE_{\gamma}\boldsymbol{\Phi}(\mathbf{q}_{\parallel},z_{0},E_{\gamma})\cdot
\mathbf{a}(\mathbf{q}_{\parallel},z_{max},E_{\gamma}),\label{eq:jsc_abs}
\end{align}
where the absorptance corresponds to the external quantum efficiency
$EQE(\mathbf{q_{\parallel}},E_{\gamma})$ in this limit. On the other hand, the short circuit current
derives from the quantities computed within the NEGF formalism as follows\cite{ae:prb_11}:
\begin{align}
J_{sc}=&j_{c}(z_{max})-j_{c}(z_{0})=\int_{z_{0}}^{z_{max}} dz ~\partial_{z}j(z)\\\equiv&
e\int_{z_{0}}^{z_{max}} dz~\mathcal{G}_{c}(z),\label{eq:jsc_gen}
\end{align}
where $j_{c}$ denotes electron current in the conduction band, which is given terms of the charge
carrier Green's functions via
\begin{align}
j_{c}(z)=\lim_{z'\rightarrow
z}\frac{e\hbar}{m_{0}}(\partial_{z}-\partial_{z'})\mathcal{A}^{-1}\sum_{\mathbf{k}_{\parallel}}
\int\frac{dE}{2\pi}G_{cc}^{<}(\mathbf{k}_{\parallel},z,z',E).
\end{align}
Together, Eqs.~\eqref{eq:jsc_abs} and \eqref{eq:jsc_gen} yield the following expression of the
absorptance in terms of the local generation spectrum:
\begin{align}
a_{\mu}(\mathbf{q}_{\parallel},z_{max},E_{\gamma})=&\Phi^{-1}_{\mu}(\mathbf{q}_{\parallel},z_{0},E_{\gamma})\int_{z_{0}}^{z_{max}}dz~
g^{\mu}(\mathbf{q}_{\parallel},z,E_{\gamma}).
\label{eq:absorpt_gen}
\end{align}

\subsubsection{Coupling to classical fields}

For classical fields, the local photon flux is given in terms of the modal components of the Poynting vector $s$ and the
electromagnetic vector potential in the following way:
\begin{align}
\Phi_{\mu}(\mathbf{q}_{\parallel},z,E_{\gamma})=&s^{\mu}(\mathbf{q}_{\parallel},z,E_{\gamma})/E_{\gamma}\\=&
2n_{r}(\mathbf{q}_{\parallel},z,E_{\gamma})c_{0}\varepsilon_{0}\hbar^{-2}E_{\gamma}|A_{\mu}(\mathbf{q}_{\parallel},z,E_{\gamma})|^{2},
\end{align}
where $n_{r}$ is the local refractive index. Using the electron-photon self-energy
\eqref{eq:se_coh} in slab representation, i.e.,
\begin{align}
&\Sigma_{cc}^{<}(\mathbf{k}_{\parallel},z,z',E)=\Big(\frac{e}{m_{0}}\Big)^{2}
\sum_{\mu\nu}p_{cv}^{\mu}(z)p_{cv}^{\nu*}(z')\nonumber\\&\times\mathcal{A}^{-1}\sum_{\mathbf{q}_{\parallel}}\int dE_{\gamma}\Big[
A_{\mu}(\mathbf{q}_{\parallel},z,E_{\gamma})A_{\nu}^{*}(\mathbf{q}_{\parallel},z',E_{\gamma})\nonumber\\&\qquad\qquad\qquad\qquad\times\tilde{G}_{vv}^{<}(\mathbf{k}_{\parallel}-\mathbf{q}_{\parallel},z,z',E-
E_{\gamma})\Big], 
\end{align}
in expression \eqref{eq:integral_genrate} for $\mathcal{G}$, the local spectral photogeneration rate acquires the following form:
\begin{align}
g^{\mu}(\mathbf{q}_{\parallel},z,E_{\gamma})=&\frac{i}{\hbar\mu_{0}}
A_{\mu}(\mathbf{q}_{\parallel},z,E_{\gamma})\int dz'
A^{*}_{\mu}(\mathbf{q}_{\parallel},z',E_{\gamma})\nonumber\\&\times
\Pi_{\mu\mu}^{>}(\mathbf{q}_{\parallel},z',z,E_{\gamma}),\label{eq:rate_coh}
\end{align}
where $\Pi$ is the photon self-energy related to the non-equilibrium polarization function of
the semiconductor slab and the momentum matrix elements,
\begin{align}
\Pi_{\mu\nu}^{>}(\mathbf{q}_{\parallel},z,z',E_{\gamma})=&-i\hbar\mu_{0}\Big(\frac{e}{m_{0}}\Big)^{2}p_{cv}^{\mu*}(z)p_{cv}^{\nu}(z')\nonumber\\&\times
\mathcal{P}_{cv}^{>}(\mathbf{q}_{\parallel},z,z',E_{\gamma}), 
\end{align}
with the RPA interband polarization function given in terms of the charge carrier GF as follows:
\begin{align}
 \mathcal{P}_{cv}^{>}(\mathbf{q}_{\parallel},z,z',E_{\gamma})=&\mathcal{A}^{-1}\sum_{\mathbf{k}_{\parallel}}\int
 \frac{dE}{2\pi\hbar}G_{cc}^{>}(\mathbf{k}_{\parallel},z,z',E)\nonumber\\&\times
 G_{vv}^{<}(\mathbf{k}_{\parallel} -\mathbf{q}_{\parallel},z',z,E-E_{\gamma}).\label{eq:polfun}
\end{align}
The absorptance of the slab of thickness $d=z_{max}-z_{0}$ required for the optical estimate of the short circuit current via
\eqref{eq:jsc_abs} is thus given by  
\begin{align}
&a_{\mu}(\mathbf{q}_{\parallel},z_{max},E_{\gamma})=\Phi^{-1}_{\mu}(\mathbf{q}_{\parallel},z_{0},E_{\gamma})\nonumber\\&\qquad\qquad\times
\frac{i}{\hbar\mu_{0}}\int_{z_{0}}^{z_{max}}dz\int_{z_{0}}^{z_{max}}dz'\Big[
A_{\mu}(\mathbf{q}_{\parallel},z,E_{\gamma})\nonumber\\&\qquad\qquad\times
A^{*}_{\mu}(\mathbf{q}_{\parallel},z',E_{\gamma})\Pi_{\mu\mu}^{>}(\mathbf{q}_{\parallel},z',z,E_{\gamma})\Big].
\end{align}
\begin{figure}[tb]
\includegraphics[width=0.5\textwidth]{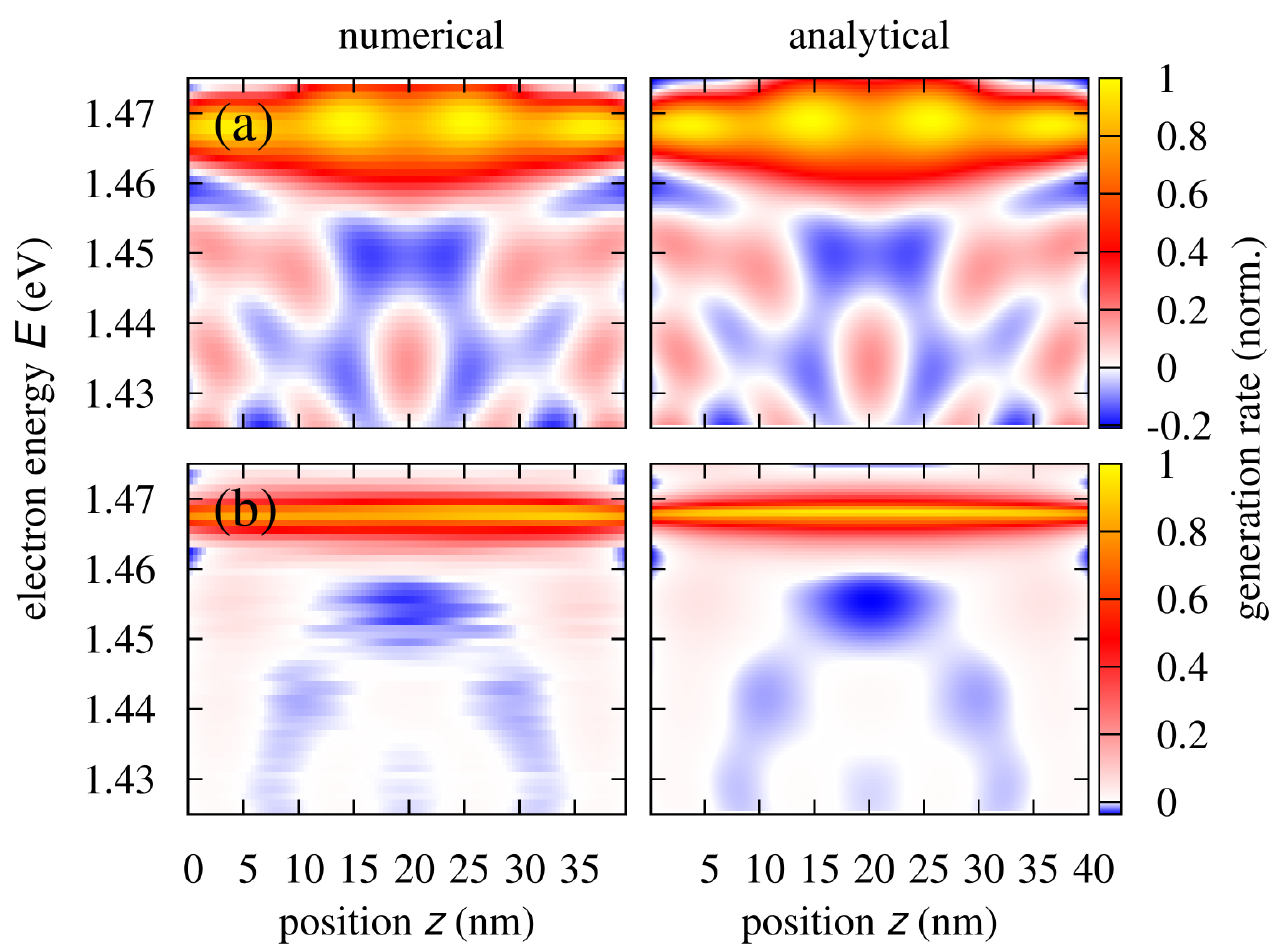}
\caption{(Color online) Local electron generation rate as given by the energy integrand of
\eqref{eq:integral_genrate} for the two band effective mass model of a homogeneous (bulk-like) GaAs slab under monochromatic
illumination with $E_{\gamma}=1.48$ eV. The rate is normalized to the local photon flux and to the
maximum value. Both the $k_{\parallel}=0$ component (a) and the momentum integrated
rate (b) show a good agreement between the simulation based on the numerical solution of the NEGF
transport problem at zero bias and the analytical evaluation of \eqref{eq:integral_genrate}
using the exact GF expressions \eqref{eq:GF_exact_in} and \eqref{eq:GF_exact_out}.
\label{fig:locrate}}
\end{figure}
 
To verify the accuracy of the numerical approach chosen to compute the microscopic non-local
response of a thin semiconductor slab, the numerical generation rate is compared to the analytical
result for the integrand of \eqref{eq:polfun} based on the exact Green's functions of homogeneous
bulk material given in the Appendix. The material parameters used are $m_{c}=0.067~m_{0}$,
$m_{v}=0.25~m_{0}$, $E_{g}=1.42$~eV $\equiv E_{c0}$,  $E_{v0}=0$~eV and
$|\bar{p}_{cv}|^{2}=25$~eV$\cdot~ m_{0}$. Fig.~\ref{fig:locrate}(a) shows the $k_{\parallel}=0$ component of the
local electron generation rate spectrum for monochromatic illuminaton normalized to the local photon flux, which then basically 	amounts to the evaluation of 
the energy integrand of the polarization function \eqref{eq:polfun} 	at photon energy
$E_{\gamma}=1.48$ eV. 	The spectral and spatial pattern are in excellent 	agreement with the
analytical 	high-resolution result. The same holds for the $\mathbf{k}_{\parallel}$-integration of 
the spectral rate, as inferred from Fig.~\ref{fig:locrate}(b). 

\begin{figure}[tb]
\includegraphics[width=0.45\textwidth]{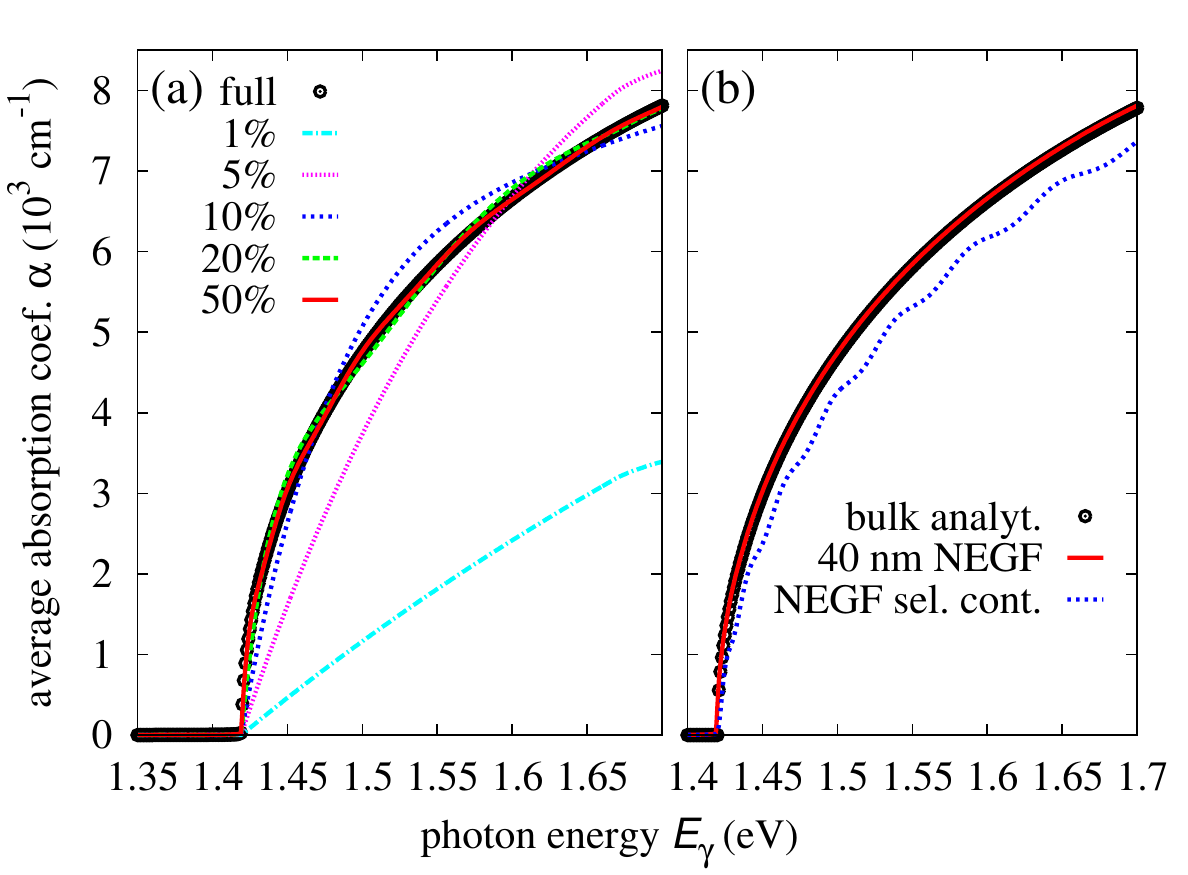}
\includegraphics[width=0.25\textwidth]{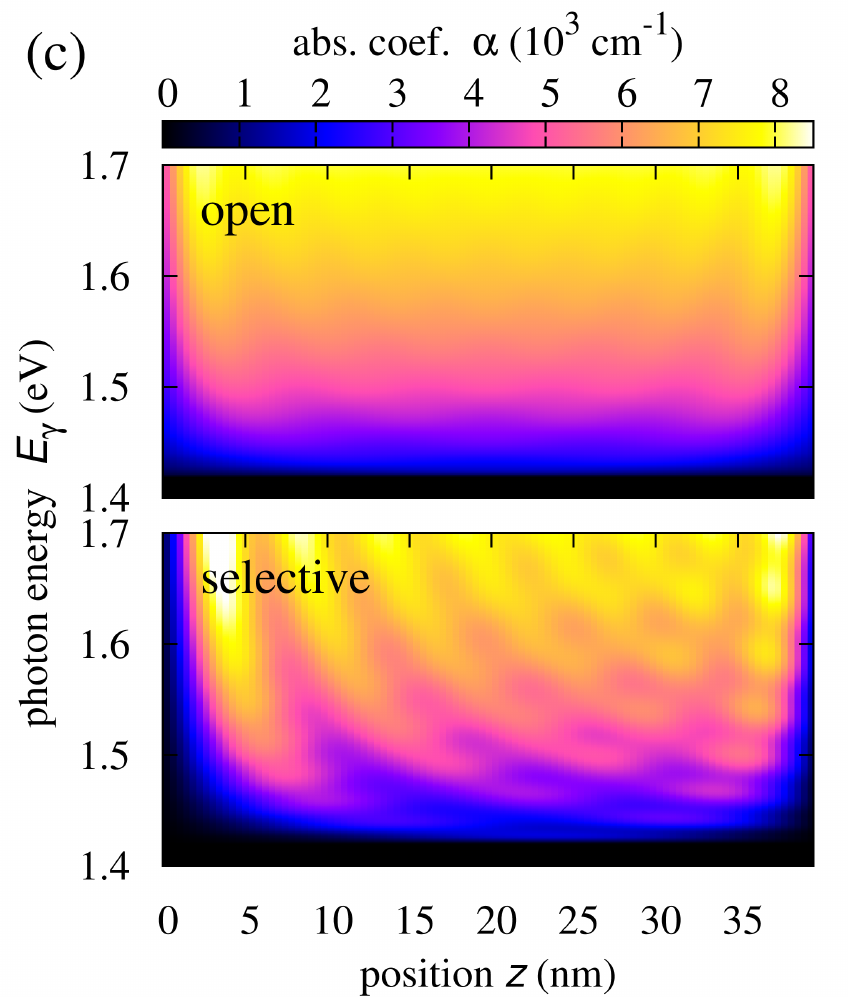}\includegraphics[width=0.23\textwidth]{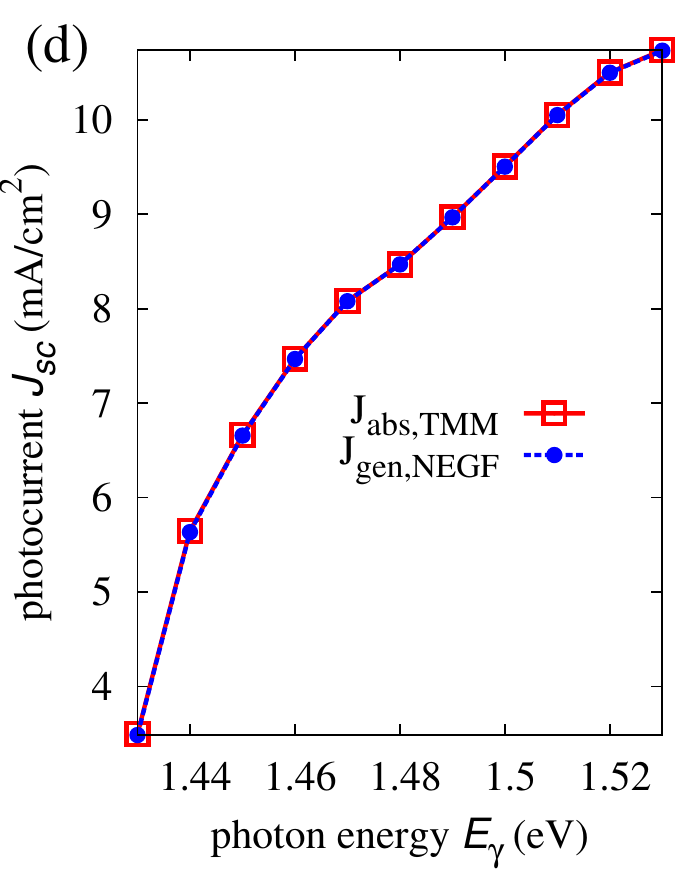}
 \caption{(Color online) (a) Spatially averaged absorption coefficient of an
electronically open 40-nm-thick GaAs slab, for consideration of different fractions of off-diagonals
in the charge carrier Green's functions. (b) The absorption coefficient from the full
matrix is in excellent agreement with the analytical result for bulk, while it is slightly reduced
and shows additional oscillatory features for selective contacts. (c) Local absorption coefficient
for open and selectively contacted slab systems, with interferences from reflections and
zero magnitude minima from wave function nodes at closed contacts. (d) Comparison of the
photocurrent as obtained from the absorptance  with the terminal current from the full NEGF
transport simulation based on the same illumination and extinction coefficient.\label{fig:abs_flat}}
\end{figure}

In most cases relevant for
optoelectronic devices, the electronic spatial correlations, i.e., the off-diagonal elements of the 
interband polarization function, decay much faster than the amplitude of the vector potential inside
the absorber, and $A(z')\approx A(z)$ can be assumed in \eqref{eq:rate_coh}. In this case, the
local absorption coefficient acquires the simple form
\begin{align}
\alpha_{\mu}(\mathbf{q}_{\parallel},z,E_{\gamma})=&\frac{\hbar c_{0}}{2
n_{r}(\mathbf{q}_{\parallel},z,E_{\gamma})E_{\gamma}}\nonumber\\&\times\int dz' 
\mathrm{Re}\Big[i\Pi_{\mu\mu}^{>}(\mathbf{q}_{\parallel},z',z,E_{\gamma})\Big] \\
=&\frac{\hbar^{2}}{2n_{r}(\mathbf{q}_{\parallel},z,E_{\gamma})c_{0}\varepsilon_{0}E_{\gamma}}
\Big(\frac{e}{m_{0}}\Big)^{2}\mathrm{Re}\Big[p_{cv}^{\mu}(z)\nonumber\\&\times
\int dz'p_{cv}^{\mu*}(z')
\mathcal{P}_{cv}^{>}(\mathbf{q}_{\parallel},z',z,E_{\gamma})\Big],\label{eq:locabscoef}
\end{align}
which is given solely in terms of the electronic structure and does not include any information on
the propagation of the light. However, the inclusion of off-diagonal elements in the electronic
Green's functions is crucial to account for the non-local nature of electron-photon
interaction \cite{pourfath:09}. Figure \ref{fig:abs_flat}(a) shows the spatially averaged local
absorption coefficient of a homogeneous GaAs slab of 40 nm thickness for different numbers of off-diagonals
considered in the electronic Green's functions: more than 20\% of the off-diagonals need to be
included for an acceptable reproduction of the full rank result. For the slab thickness chosen, the
full matrix yields already an absorption coefficient in close agreement with the analytical bulk
result for the two-band model, as seen in Fig.~\ref{fig:abs_flat}(b). The spatial resolution of the
local absorption coefficient is given in Fig.~\ref{fig:abs_flat}(c) for the bulk-like system
(``open'') 	and the slab with carrier selective contacts (``selective''). The minimum of the
absorption of 	the open system close to the contacts is due to the assumption of vanishing
off-diagonal 	contributions to the polarization function in  \eqref{eq:polfun} from outside the
slab, 	i.e., it assumed that there is no coherence between the carrier wave functions inside the
slab 	absorber and in the contact, respectively. In the system 	with selective contacts, either 
electron or hole Green's functions vanish at the contacts, causing 	nodes in the local absorption coefficient.

\begin{figure}[tb]
\includegraphics[width=0.5\textwidth]{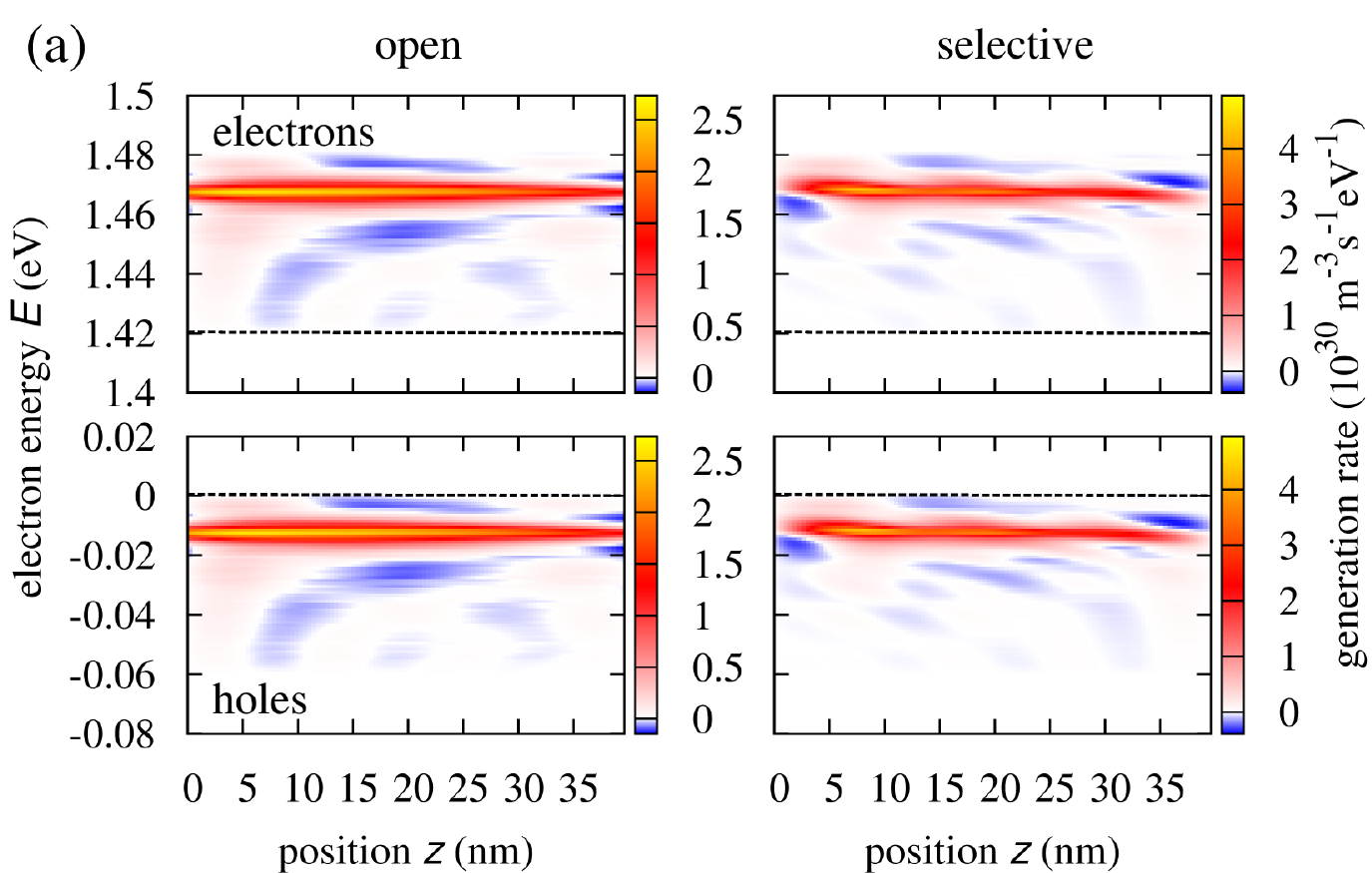}
\includegraphics[width=0.4\textwidth]{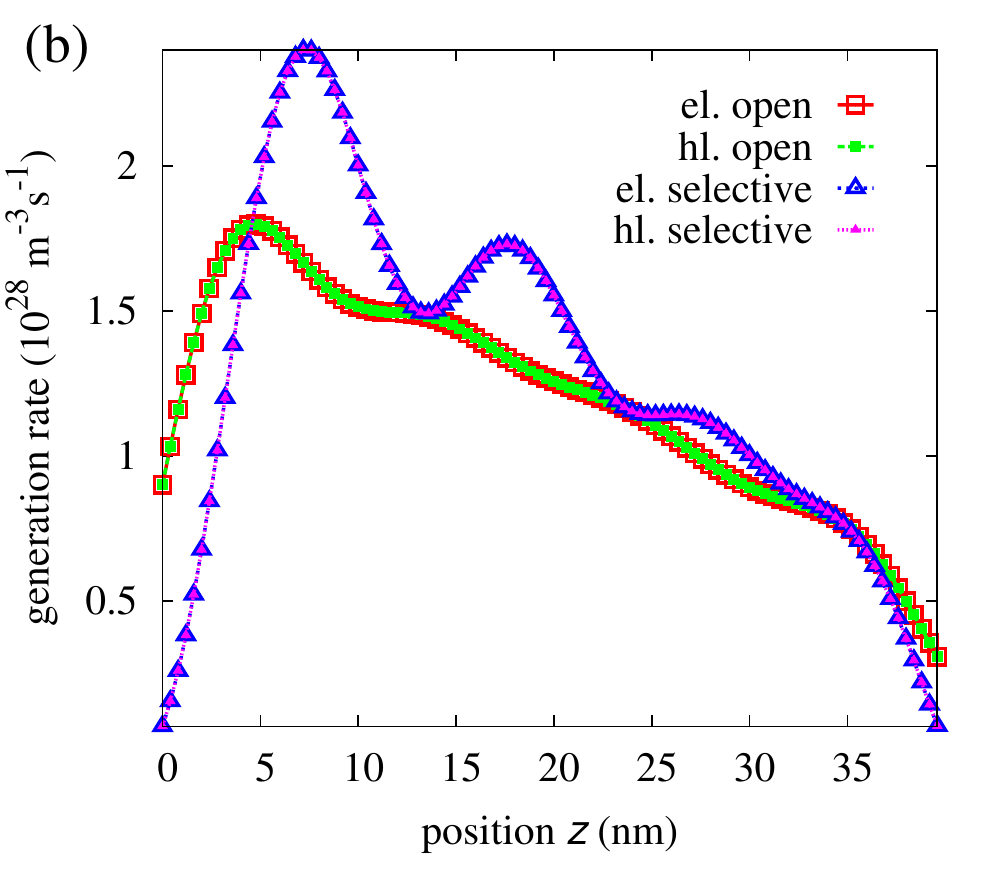}
\caption{(Color online) (a) Local generation rate spectrum for open and
selectively contacted 40 nm GaAs slab with 100 nm Ag reflector and under
monochromatic, normally incident light of 1 kW/m$^{2}$ at $E_{\gamma}=1.48$ eV; (b) Integrated local
carrier generation rate, which is identical for electrons and holes, with closed contacts resulting
in reflection-induced interference effects and boundary nodes.
\label{fig:locrate_elhl}}
\end{figure}  
\begin{figure}[t]

 \includegraphics[width=0.5\textwidth]{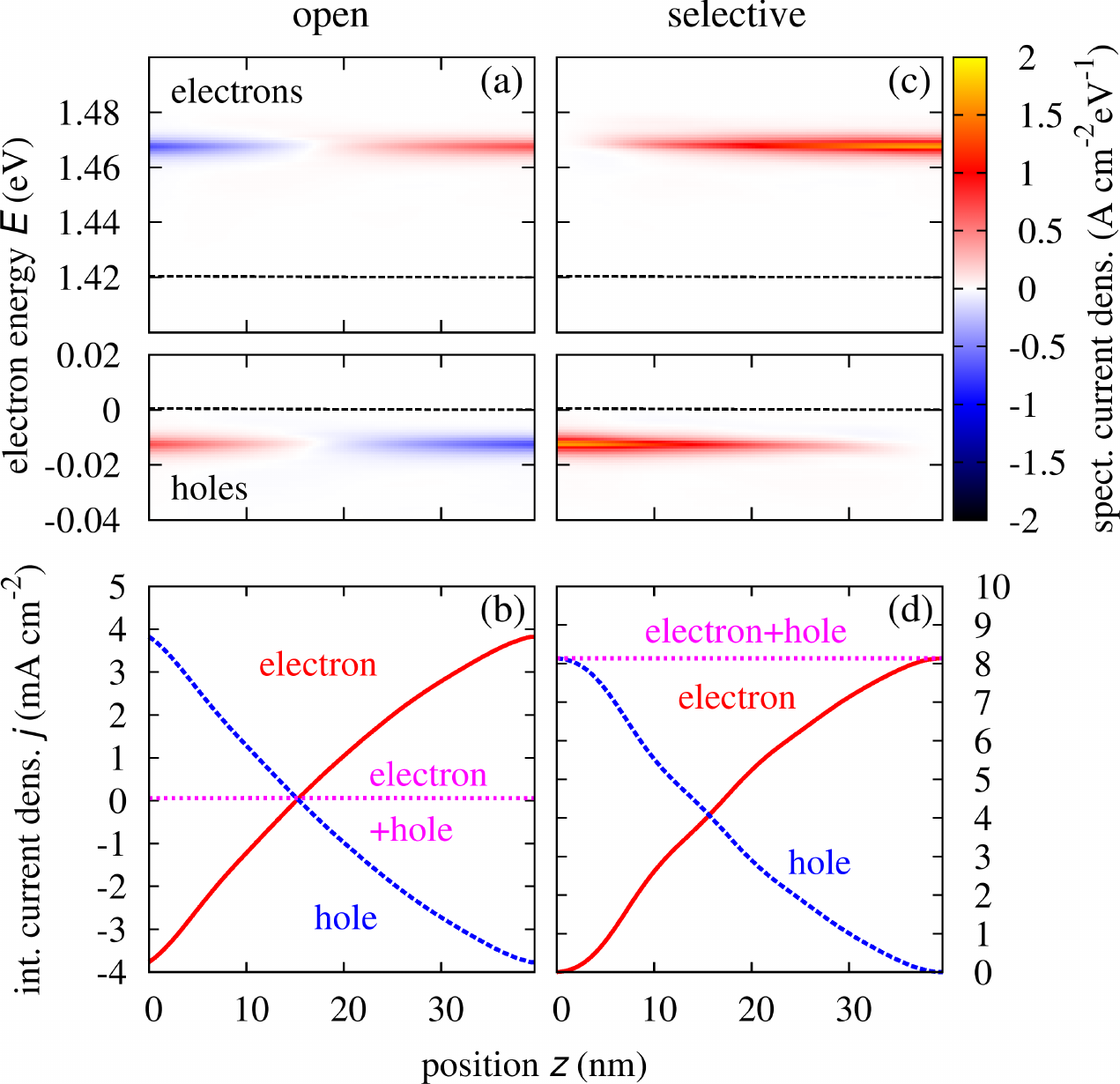}
 \caption{(Color online) (a) Spatially resolved photocurrent spectrum for the open slab under monochromatic
 illumination of 1 kW/m$^{2}$ with $E_{\gamma}=1.48$ eV; in the absence of carrier selective
 contacts, diffusion leads to inverse current flow towards the minority contacts, with the result of vanishing net
 current upon energy integration as displayed in (b). (c) For carrier selective contacts, reverse
 current components are small, and the net integral current (d) is strictly positive and perfectly
 conserved.\label{fig:loccurr_all}} 
\end{figure} 
  
The absorptance can also be computed directly from the Poynting vector based on the fields obtained from
the TMM,
 \begin{align}
 a_{\mu}(\mathbf{q}_{\parallel},z_{max},E_{\gamma})=1-s^{\mu}(\mathbf{q}_{\parallel},z_{max},E_{\gamma})
 /s^{\mu}(\mathbf{q}_{\parallel},z_{0},E_{\gamma}),
 \end{align}
for flux incident at $z=z_{0}$. Figure \ref{fig:abs_flat}(c) shows the close agreement between the
monochromatic photocurrent from the absorptance as given by the integrand of \eqref{eq:jsc_abs} and
the terminal current obtained from the NEGF for the same monochromatic illumination, using in the
TMM an extinction coefficient computed from the electronic structure in consistence with $a$ via
\begin{align}
\kappa_{\mu}(\mathbf{q}_{\parallel},z,E_{\gamma})=&\alpha_{\mu}(\mathbf{q}_{\parallel},z,E_{\gamma})\cdot\frac{\hbar
c_{0}}{2E_{\gamma}}\\
=&\frac{(\hbar c_{0})^{2}}{4
n_{r}(\mathbf{q}_{\parallel},z,E_{\gamma})E_{\gamma}^{2}}\nonumber\\&\times\int dz'
\mathrm{Re}\Big[i\Pi_{\mu\mu}^{>}(\mathbf{q}_{\parallel},z',z,E_{\gamma})\Big]\label{eq:extcoef}
\end{align}
 and for a 100 nm Ag back reflector ($n_{r}=0.16$, $\kappa=5.85$).

The proper generation rate under consideration of the spatial variation of
the electromagnetic field inside the slab is shown in Fig.~\ref{fig:locrate_elhl} for both the 
bulk-like slab and the system with carrier-selective contacts under 
monochromatic illumination at $E_{\gamma}=1.48$ eV, 1~kW/m$^{2}$ and normal incidence. The spectral
generation patterns shown in Fig.~\ref{fig:locrate_elhl}(a) are identical for electrons and holes, and show
weak negative features 	away from the resonance. However, the energy-integrated local generation
rate, 	displayed in Fig.~\ref{fig:locrate_elhl}(b) is strictly positive. Again, the imposition of 
carrier-selective contacts 	modifies the local absorption due to additional interferences from 
reflections and magnitude zeros from wave function nodes at closed contacts.

The local photocurrent spectrum induced in the open bulk-like slab system by the local carrier
generation rate is displayed in Fig.~\ref{fig:loccurr_all}(a). In the absence of carrier selective
contacts, carriers diffuse symmetrically in both directions, with the result of vanishing net
integral current [Fig.~\ref{fig:loccurr_all}(b)]. For the selectively contacted system, the
negative contributions in the generation rate give rise to reverse flow at certain energies 
(this was 	also observed for purely 1D systems in Ref.~\onlinecite{buin:13}), as revealed in
Fig.~\ref{fig:loccurr_all}(c), however, like in the case of the generation rate,  the
observable integral current is always positive and the sum of electron and hole
current contributions is perfectly conserved, as shown in Fig.
\ref{fig:loccurr_all}(d).

\subsubsection{Coupling to the photon GF}  
Using the slab-expression for the steady-state RPA electron-photon self-energy,
\begin{align}
&\Sigma^{\lessgtr}(\mathbf{k}_{\parallel},z,z',E)=i\hbar\mu_{0}\Big(\frac{e}{m_{0}}\Big)^{2}
\sum_{\mu\nu} \hat{p}^{\mu}(z)
\hat{p}^{\nu*}(z')\nonumber\\&\times\mathcal{A}^{-1}\sum_{\mathbf{q}_{\parallel}}\int
\frac{dE_{\gamma}}{2\pi\hbar}\Big[
G^{\lessgtr}(\mathbf{k}_{\parallel}-\mathbf{q}_{\parallel},z,z',E-E_{\gamma})\nonumber\\&\hspace{3cm}\times
\mathcal{D}_{\mu\nu}^{\lessgtr}(\mathbf{q}_{\parallel},z,z',E_{\gamma})\Big],
\label{eq:elphotse_rpa_slab}
\end{align}
in Eq.~\eqref{eq:integral_genrate}, the local modal generation rate acquires the following form:
\begin{align}
g^{\mu}(\mathbf{q}_{\parallel},z,E_{\gamma})=&-(2\pi\hbar)^{-1}\sum_{\nu}\int
dz'\Big[\mathcal{D}_{\mu\nu}^{<}(\mathbf{q}_{\parallel},z,z',E_{\gamma})
\nonumber\\&\quad\times\Pi_{\nu\mu}^{>}(\mathbf{q}_{\parallel},z',z,E_{\gamma})\Big].\label{eq:rate_inc}
\end{align} 
In terms of the photon slab Green's functions defined in
\eqref{eq:photongf_slab}, the contribution of the $\mu$-polarization to the $z$-component of the spectral Poynting vector reads
\begin{align}
s^{\mu}_{z}(\mathbf{q}_{\parallel},z,E_{\gamma})=&-\frac{E_{\gamma}}{2\pi\hbar}\lim_{z'\rightarrow
z}\partial_{z'}\mathrm{Re}\Big[\mathcal{D}_{\mu\mu}^{>}(\mathbf{q}_{\parallel},z,z',E_{\gamma})\nonumber\\&+
\mathcal{D}_{\mu\mu}^{<}(\mathbf{q}_{\parallel},z,z',E_{\gamma})\Big].
\end{align}  
\begin{figure}[t]
\begin{center}
 \includegraphics[width=0.35\textwidth]{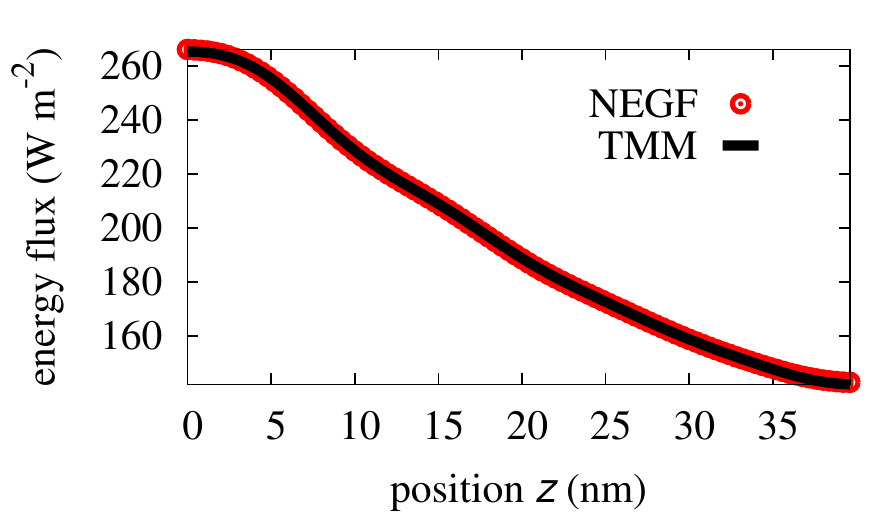}
 \caption{(Color online) Photon energy flux in the selectively contacted slab as computed via TMM
 and NEGF methods, showing the close agreement of the two approaches for coherent light propagation, i.e., in absence
 of spontaneous emission.\label{fig:flux}}
 \end{center}
\end{figure}  
\begin{figure}[t]
 \includegraphics[width=0.5\textwidth]{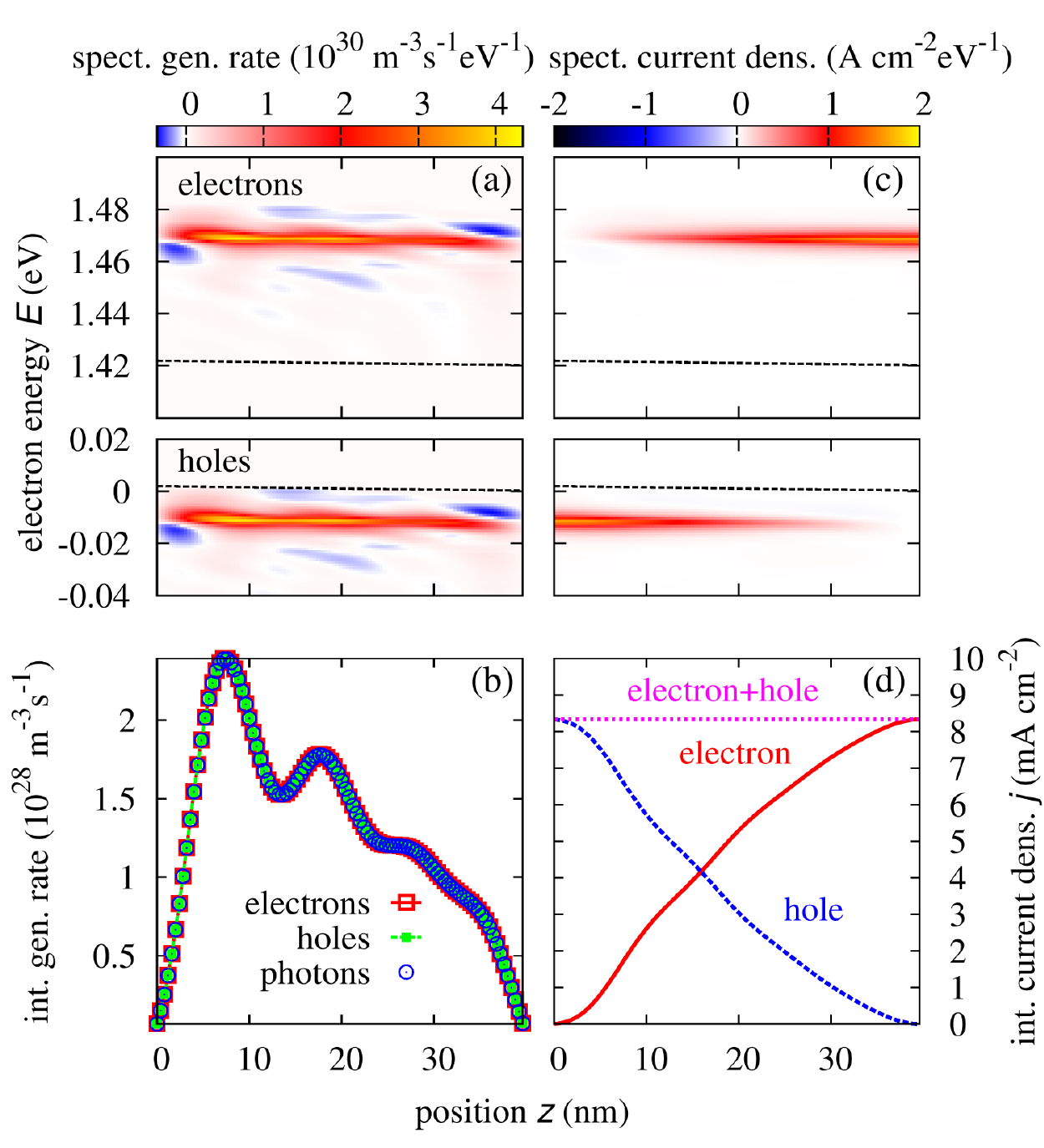}
 \caption{(color online) (a) Local charge carrier generation rate spectrum, (b) local carrier
 generation rate and optical rate, (c) local charge carrier current spectrum and (d) integral current for the
 selectively contacted slab system, as provided by the NEGF formalism using the coupling to the
 photon GF. The results are in close agreement with those from the coherent coupling
 approximation. The local charge carrier generation rate concides exactly with the optical
 rate from photon self-energy and GF.
 \label{fig:loc_curr_rate}} 
\end{figure}
Figure \ref{fig:flux} shows the close agreement of the optical energy flux as computed via TMM and the
photon NEGF formalism with fully non-local photon self-energy $\Pi$\cite{ae:photgf}. From the modal
terms of local rate and photon flux, the local absorption coefficient at given angle of incidence
and polarization is then given by \eqref{eq:genrate_def} with 
$\Phi_{\mu}(\mathbf{q}_{\parallel},z,E_{\gamma})=s^{\mu}_{z}(\mathbf{q}_{\parallel},z,E_{\gamma})/E_{\gamma}$,
and from \eqref{eq:absorpt_gen}, the absorptance follows as
\begin{align}
&a_{\mu}(\mathbf{q}_{\parallel},z_{max},E_{\gamma})=-\big[2\pi\hbar~\Phi_{\mu}
(\mathbf{q}_{\parallel},z_{0},E_{\gamma})\big]^{-1}\nonumber\\&\hspace{1cm}\times 
\int_{z_{0}}^{z_{max}}dz\int_{z_{0}}^{z_{max}}
dz'\sum_{\nu}\Big[\mathcal{D}_{\mu\nu}^{<}(\mathbf{q}_{\parallel},z,z',E_{\gamma})\nonumber\\&\hspace{1cm}\times
\Pi_{\nu\mu}^{>}(\mathbf{q}_{\parallel},z',z,E_{\gamma})\Big].
\end{align}
It should be noted that the above expression for the absorptance does not consider any increase in
the photon flux due to emission processes; to that end, the self-energy component $\Pi^{>}$ needs to
be replaced by $\hat{\Pi}=\Pi^{>}-\Pi^{<}$ \cite{richter:08}, which for the present case of
short circuit conditions is virtually identical to $\Pi^{>}$. If reabsorption is neglected, the GF component $\mathcal{D}^{<}$ is directly proportional to the photon flux via $\mathcal{D}_{0}^{<}=\hat{\mathcal{D}}_{0}\cdot \Phi_{0}$, where $\hat{\mathcal{D}}_{0}$ is independent from the excitation due to the photon flux $\Phi_{0}$ incident at $z=z_{0}$\cite{ae:photgf}. With that, the absorptance acquires the form  
\begin{align}
&a_{\mu}(\mathbf{q}_{\parallel},z_{max},E_{\gamma})=-(2\pi\hbar)^{-1}\nonumber\\&\hspace{1cm}\times 
\int_{z_{0}}^{z_{max}}dz\int_{z_{0}}^{z_{max}}
dz'\sum_{\nu}\Big[\hat{\mathcal{D}}_{\mu\nu}(\mathbf{q}_{\parallel},z,z',E_{\gamma})\nonumber\\&\hspace{1cm}\times
\hat{\Pi}_{\nu\mu}(\mathbf{q}_{\parallel},z',z,E_{\gamma})\Big],
\end{align}
where $\hat{\boldsymbol{\mathcal{D}}}=\boldsymbol{\mathcal{D}}^{R}\boldsymbol{
\mathcal{D}}^{R,-1}_{0}
\hat{\boldsymbol{\mathcal{D}}}_{0}\boldsymbol{\mathcal{D}}^{A,-1}_{0}
\boldsymbol{\mathcal{D}}^{A}$. This corresponds to the result found in Ref.~\onlinecite{richter:08}. 

The numerical results for local generation rate and charge carrier flow in the selectively contacted
slab for coupling to the photon GF are displayed in Fig.~\ref{fig:loc_curr_rate}. As can be inferred from
Figs.~\ref{fig:loc_curr_rate}(a) and \ref{fig:loc_curr_rate}(b) in comparison with Fig.~\ref{fig:locrate_elhl}, the values
of spectral and integral charge carrier generation rates agree closely with those provided by the
coherent coupling. The same holds for the spectral and integral currents displayed in
Figs.~\ref{fig:loc_curr_rate}(c) and \ref{fig:loc_curr_rate}(d), if compared to the results in Fig.~\ref{fig:loccurr_all}.
Furthermore, as also shown in Fig.~\ref{fig:loc_curr_rate}(b), the charge carrier generation rate 
coincides exactly with the optical rate as computed using \eqref{eq:rate_inc}, i.e., in terms of
the photon GF and self-energy. 

The coincidence with the coherent coupling approximation originates in the absence of (optical) coherence
breaking mechanisms in the situation under consideration, direct photogeneration being a stimulated
process. The photon GF can thus be directly related to the average
fields\cite{henneberger:96,richter:08}. This may be compared to the case of electron transport in
mesoscopic systems, where in absence of phase breaking, i.e., incoherent scattering mechanisms, the
Landauer formalism based on a transmission function obtained from an electronic version of the TMM
is equivalent to the NEGF picture of electron transport \cite{datta:95}.

\section{Conclusions} 
In this paper, a description of charge carrier photogeneration in thin semiconductor films within
the NEGF formalism was established. An effective electron-photon self-energy was derived for coherent coupling to
classical fields. Numerical simulations were performed for a thin, selectively contacted semiconductor slab with a back
reflector. The results for charge carrier generation rate and photocurrent computed using the NEGF formalism of carrier
transport coupled to the TMM for the electromagnetic fields are in close agreement with the predictions of both the optical estimate 
via the average absorption coefficient and the full coupled NEGF solution for the propagation of interacting charge
carriers and photons. Thus, the NEGF framework presented here is consistent with the
classical picture of light-matter coupling in the limit of optically coherent processes, and, at the same time,
enables the consideration of both extraction and radiative recombination of charge carriers in the presence of complex
nanostructure potentials. This unique capability turns the present approach into a powerful instrument for the 
investigation of the radiative efficiency limit in nanostructure-based solar cell devices.

For a quantitative analysis of photocurrent generation in realistic device structures, in addition to the use of an accurate description of the electronic structure, extension of the formalism to radiative intraband mechanisms (e.g., free carrier absorption) and non-radiative intra- and interband scattering processes (e.g., electron-phonon and Auger) will be required. The inclusion of such processes, however, while it modifies the GF and self-energies of charge carriers and photons, it does not affect the validity of the general expressions derived here for absorptance, generation rate and photocurrent in terms of the GF.  
   
\acknowledgments   
The author would like to acknowledge the support and kind hospitality of the National Renewable
Energy Laboratory in Golden, Colorado, USA, during his visit in the framework of the
Helmholtz-NREL Joint Research Initiative HNSEI. 
\newline

\appendix
 
\section{Electronic Green's functions for homogeneous slab\label{sec:appa}}
For a homogeneous bulk-like system, the effective mass approximation of the slab representation of
the steady-state Green's function components for non-interacting charge carriers in quasi-equilibrium conditions
characterized by a quasi-Fermi level $\mu$ takes the following form:
\begin{align}
G_{b0}^{<}(\mathbf{k}_{\parallel},z,z',E)=&if_{\mu_{b}}(E)A_{b0}(\mathbf{k}_{\parallel},z,z',E),\label{eq:GF_exact_in}
\\
G_{b0}^{>}(\mathbf{k}_{\parallel},z,z',E)=&-i[1-f_{\mu}(E)]A_{b0}(\mathbf{k}_{\parallel},z,z',E),\label{eq:GF_exact_out}\\
A_{b0}(\mathbf{k}_{\parallel},z,z',E)=&\frac{2
m_{b}^{*}}{\hbar^{2}}\frac{\cos[k_{z}^{b}(\mathbf{k}_{\parallel},E)(z-z')]}{k_{z}^{b}(\mathbf{k}_{\parallel},E)},
\end{align}
where
\begin{align}
k_{z}^{b}(\mathbf{k}_{\parallel},E)=&\sqrt{2 m^{*}_{b}E-\hbar^{2}k_{\parallel}^2}/\hbar,\quad
b=c,v,\\
f_{\mu}(E)=&\big\{\exp[(E-\mu)/k_{B}T]+1\big\}^{-1}.
\end{align}  

%
%
%

\bibliographystyle{apsrev4-1}
\bibliography{/home/aeberurs/Biblio/bib_files/negf,/home/aeberurs/Biblio/bib_files/aeberurs,/home/aeberurs/Biblio/bib_files/generation,/home/aeberurs/Biblio/bib_files/scqmoptics,/home/aeberurs/Biblio/bib_files/pv}

\end{document}